\documentclass[prb,twocolumn,superscriptaddress,showpacs]{revtex4}
\usepackage{graphicx}
\usepackage{amsmath}
\usepackage{amssymb}
\usepackage{color}

\renewcommand{\k}{\mathbf{k}}

\newcommand{\Tr}{\mathrm{Tr}}

\begin{document}

\title{The Tower of States and the Entanglement Spectrum in a Coplanar Antiferromagnet}
\author{Louk Rademaker}
\affiliation{Kavli Institute for Theoretical Physics, University of California Santa Barbara, CA 93106, USA}
\date{\today}

\begin{abstract} 
We extend the analytical arguments of Metlitski and Grover (arXiv:1112.5166) to compute the entanglement spectrum and entanglement entropy of coplanar antiferromagnets with $SO(3)$ order parameter symmetry. The low-energy states in the entanglement spectrum exhibit the tower of states structure, as is expected for systems that undergo spontaneous symmetry breaking. Our results are consistent with numerical results on the triangular and Kagom\'{e} lattice.
\end{abstract}

\pacs{05.30.-d,75.30.-m}

\maketitle

A system that exhibits spontaneous continuous symmetry breaking (SSB) in the thermodynamic limit $N \rightarrow\infty$, where the order parameter does not commute with the Hamiltonian, will for any finite system size have a unique ground state. For example, a finite size quantum antiferromagnet will have a completely symmetric spin singlet as a ground state.

How then can any finite system show signatures of its symmetry breaking fate? Anderson\cite{Anderson:InL7mPjj,Anderson:2011vu} pointed out that such a finite size system contains a 'tower of states' or 'thin spectrum'\cite{VanWezel:liWTmDzL,Lhuillier:2005us}: eigenstates with an energy $\mathcal{O}(1/N)$ that vanishes in the thermodynamic limit, and who have a degeneracy structure that reveals the order parameter symmetry. This tower of states can be used to discover the symmetry broken state numerically\cite{Azaria93,BBernu:2011te}, which deals, by its very nature, with finite systems only.

Recently, the notion that the ground state entanglement can reveal the low-energy spectrum of a phase has gained widespread recognition.\cite{Li:2008cg} For SSB-systems, this suggests that information about the symmetry breaking lies hidden in the completely symmetric ground state. Indeed, it has been shown numerically for the collinear antiferromagnet that the entanglement entropy does exhibit the tower of states by having an extra logarithmic term.\cite{Kallin2011PhRvB..84p5134K} These results were confirmed analytically using the non-linear sigma model.\cite{Metlitski:akSbyvLD}

In this paper we extend the analytical results of Ref. [\onlinecite{Metlitski:akSbyvLD}] to explain the numerical results\cite{Kolley2014arXiv1410.7911K,Kolley2013arXiv1307.6592K} for the entanglement structure of \emph{coplanar antiferromagnets} on the triangular and Kagom\'{e} lattice. Unlike collinear antiferromagnets, the order parameter now has an $SO(3)$ symmetry, which changes the degeneracies of the tower of states. In Sec. \ref{SecAFM} we will introduce the concept of $SO(3)$ order, and the corresponding tower of states. We will then show in Sec. \ref{SecEntg} that, consistent with the numerical results, the tower of states will also be present in the entanglement spectrum of the ground state. Our results support the thesis that symmetry broken systems can be understood by studying their finite size unique symmetric ground state.

\section{Coplanar antiferromagnets}
\label{SecAFM}

In this section we review the main relevant properties of coplanar antiferromagnets, which will lead up to the representation of coplanar systems in terms of an $SO(3)$ nonlinear sigma model.

\subsection{Triangular antiferromagnet}
\begin{figure}
 \includegraphics[width=\columnwidth]{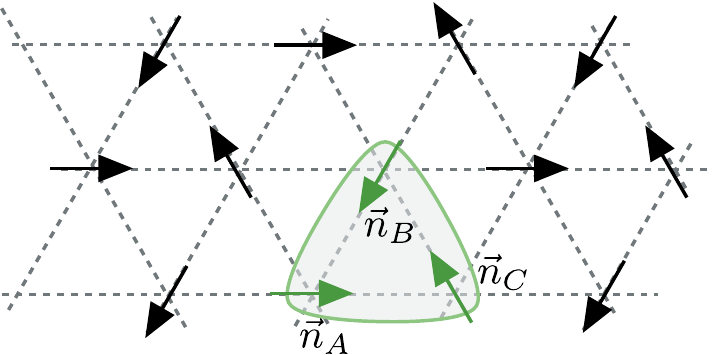}
 \caption{\label{FigTriangular} (color online) 
The triangular lattice with the classical 120 degree coplanar ordered state. This configuration is used as reference state, see Eqn. (\ref{RefV3}). }
\end{figure}

As a prime example of a coplanar antiferromagnet, let us consider the  Heisenberg model
\begin{equation}
	H = J \sum_{\langle ij \rangle} \vec{S}_i \cdot \vec{S}_j,
	\label{Heisenberg}
\end{equation}
with $J>0$ and the sum over nearest neighbor sites, on a \emph{triangular lattice}. On a triangular lattice, the unit vectors that connect neighboring sites are
\begin{eqnarray}
	\vec{e}^1 &=& (1,0), \nonumber \\
	\vec{e}^2 &=& (-\frac{1}{2}, \frac{1}{2} \sqrt{3} ), \nonumber\\
	\vec{e}^3 &=& (-\frac{1}{2}, -\frac{1}{2} \sqrt{3} ),\label{VectorsTriangular3}
\end{eqnarray}	
Unlike its square lattice companion, the triangular antiferromagnet is frustrated already at the classical level. Nonetheless, a classical ground state exists where the triangular lattice is split into \emph{three} sublattices A, B and C, and spins on different sublattices make a 120 degree angle which each other. For example, we choose an explicit reference state (see Fig. \ref{FigTriangular}),
\begin{eqnarray}
	\vec{n}_A &=& ( 1,0,0), \nonumber \\
	\vec{n}_B &=& (-\frac{1}{2}, -\frac{1}{2} \sqrt{3}, 0), \nonumber \\
	\vec{n}_C &=& (-\frac{1}{2}, \frac{1}{2} \sqrt{3}, 0), \label{RefV3}
\end{eqnarray}
such that our reference ordered state will have $\langle \vec{S}_{i \in A} \rangle = \sigma \vec{n}_A$, etcetera, where $\sigma$ is the spin of a single site.

The order parameter is an element of the rotation group $SO(3)$: each ordered state can be obtained from the reference state of Eqn. (\ref{RefV3}) by a unique rotation $\hat{R} \in SO(3)$ acting on the reference state. This makes the triangular antiferromagnet \emph{coplanar}, in contrast with the square lattice antiferromagnet which is \emph{collinear} and where the order parameter is just a vector lying on the $2$-sphere.

The question is whether the ground state of the \emph{quantum} triangular antiferromagnet bears any resemblance to the classical 120 degree state. As early as 1973 Anderson proposed that the ground state of Eqn. (\ref{Heisenberg}) is not ordered, but instead a resonating valence bond 'spin liquid' state.\cite{Anderson:1973eo} Contradicting numerical results (for example Ref. [\onlinecite{Suzuki:2014gj}] versus Ref. [\onlinecite{BBernu:2011te}]) keep this an open problem, but for the present purpose, we assume that the model does exhibit symmetry breaking. To avoid the subtleties associated with a possible spin liquid phase, one can stabilize the 120-degree ordered state by adding a small ferromagnetic coupling between spins on the same sublattice.

Similarly, coplanar order can be found in Kagom\'{e} antiferromagnets.

\subsection{Tower of States}

On any finite size system, however, the ground state of the triangular antiferromagnet has not broken the spin rotation symmetry. After all, the total spin $\vec{S} = \sum_i \vec{S}_i$ commutes with the Hamiltonian and thus the ground state is an eigenstate of total spin too. It turns out the ground state is given by the symmetric singlet ($S=0$) state.\footnote{Marshall\cite{1955RSPSA.232...48M} proved that the ground state of the Heisenberg model on any bipartite lattice must be a singlet. To our knowledge such a proof has not been provided for the triangular lattice, though it is usually assumed that it is a spin singlet and numerical computations corroborate that.\cite{Kolley2013arXiv1307.6592K}} 

Insights into this singlet ground state and its low energy excitations can be gained from separating the Heisenberg Hamiltonian into a spin-wave part and terms that depend on the total spin on each sublattice.\cite{VanWezel:liWTmDzL,Anderson:InL7mPjj,Anderson:2011vu,Azaria93,BBernu:2011te,Lhuillier:2005us} To separate the short-wavelength from long-wavelength physics, we perform a Fourier transform $\vec{S}_i = \frac{1}{\sqrt{N}} \sum_k e^{ikr_i} \vec{S}_k$, 
\begin{eqnarray}
	H = \frac{1}{2} J \sum_{k} \gamma_\k \vec{S}_{\k} \cdot \vec{S}_{-\k}
	\label{Hmomentum}
\end{eqnarray}
where $\gamma_\k$ is a geometric factor associated with the lattice: $\gamma_\k = \sum_{\delta} e^{ik\delta}$ where $\delta$ are the four six neighbors on the triangular lattice as defined in Eqn. (\ref{VectorsTriangular3}). At the momenta $\k=(0,0)$, $\k = \mathbf{K} = (2\pi, 2\pi/\sqrt{3})$ and $\k = -\mathbf{K}$ the geometric factor $|\gamma_\k|$ reaches the maximum value of $1$ which prevents a spin wave treatment at these momentum points. Isolating this part of the Hamiltonian,
\begin{equation}
	H_0 = 3J \left( \vec{S}_{\k=0}^2 - \vec{S}_{\mathbf{K}} \cdot \vec{S}_{-\mathbf{K}} \right)
\end{equation}
and expressing it in terms of the total spin $\vec{S} = \sum_{i} \vec{S}_i$ and the spin on each of the sublattices, $\vec{S}_A = \sum_{i \in A} \vec{S}_i$, etc., we uncover the triangular equivalent of the Lieb-Mattis\cite{Lieb1962JMP.....3..749L} Hamiltonian,
\begin{eqnarray}
	H_0 &=& \frac{18J}{N} \left( \vec{S}_A \cdot \vec{S}_B + \vec{S}_B \cdot \vec{S}_C + \vec{S}_A \cdot \vec{S}_C \right) \nonumber \\
		&=& \frac{9J}{N} \left( \vec{S}^2 - \vec{S}_A^2 - \vec{S}_B^2 - \vec{S}_C^2 \right).
	\label{LMtriangle}
\end{eqnarray}
This Hamiltonian can be solved exactly.\cite{Azaria93,BBernu:2011te} Its ground state is such that the spin on each sublattice is maximal and equal to $\sigma N/3$ where $N$ is the total number of sites of the system, and the total spin of the system is zero: the aforementioned singlet state. The excited states have a higher value of the total spin,
\begin{equation}
	E = \frac{9J}{N} S (S+1) + E_0.
	\label{TriangleTOS}
\end{equation}
These states are referred to as the 'tower of states' or the 'thin spectrum'.\cite{VanWezel:liWTmDzL,Anderson:InL7mPjj,Anderson:2011vu,Azaria93,BBernu:2011te,Lhuillier:2005us} The lowest energy excitation is to the total triplet state, with energy $\mathcal{O}(1/N)$ above the ground state. The eigenstates with energy Eqn. (\ref{TriangleTOS}) are $(2S+1)^2$ degenerate, since the states consist of combinations of the three sublattice spins.\cite{Azaria93,BBernu:2011te}

In the thermodynamic limit, the tower of states becomes degenerate with the singlet ground state and the symmetry broken state can be formed as a superposition of the tower states.

The remaining part of the Heisenberg Hamiltonian contains short-wavelength physics at momenta $\k \neq 0, \mathbf{K}$. It contains linearly dispersing spin waves following Goldstone's theorem. Since $\omega = c |\k|$ for small but nonzero momenta, and the lowest possible momentum on a finite $N = L \times L$ size system is $k_{\mathrm{min}} = \frac{2\pi}{L}$, the lowest state with an excited Goldstone mode in $d=2$ has an energy $\mathcal{O}(1/\sqrt{N})$ above the ground state. Therefore the tower states are energetically separated from the spin-wave states.

This separation allows for numerical detection of the tower of states, which was done by Bernu et al.\cite{BBernu:2011te} for lattices up to 36 sites big. They indeed found that the low energy states displayed the structure of Eqn. (\ref{LMtriangle}) - thus enabling to see the precursor states of spontaneous symmetry breaking in a finite size system.

\subsection{Nonlinear sigma model}

The Heisenberg antiferromagnet is amenable to a semi-classical treatment, in which the Hamiltonian is mapped onto the action of the so-called \emph{nonlinear sigma model} using spin coherent states. An introduction to this mapping for square lattice antiferromagnets is given in Refs. [\onlinecite{Sachdev2011qpt..book}] and [\onlinecite{Auerbach:1994vu}]. 

For the triangular antiferromagnet, the mapping has been first derived by Dombre and Read\cite{Dombre1989PhRvB}. Let us quickly summarize their result. The spins on each site are parametrized by coherent states $| \vec{N} \rangle$, with the property that they are eigenstates of $\vec{S}_i$ with eigenvalue the vector $\vec{N}$. In imaginary time, the action becomes
\begin{eqnarray}
	S &=& i \sigma \int_0^\beta d\tau \int_0^1 du \sum_i \vec{N}_i \cdot 
		\left( \frac{\partial N_i}{\partial u} \times \frac{\partial N_i}{\partial \tau} \right)
	\nonumber \\ &&
		+ \sigma^2 J \int_0^\beta \sum_{\langle i j \rangle} \vec{N}_i \cdot \vec{N}_j.
\end{eqnarray}
For the anticipated 120-degree coplanar order, we parametrize these vectors as
\begin{equation}
	\vec{N}_i (\tau) = \frac{\hat{R}_i(\tau) \left( \vec{n}_i + a \vec{L}_i(\tau) \right) }{
		\sqrt{1 + 2 a \vec{L}_i (\tau) \cdot \vec{n}_i + a^2 \vec{L}_i^2(\tau) }}
\end{equation}
where $\vec{n}_i$ are the reference vectors Eqn. (\ref{RefV3}), that only depend on the sublattice. The rotation matrix $\hat{R}_i(\tau)$ is an element of $SO(3)$, and $\vec{L}_i(\tau)$ expresses the average magnetization, and $a$ is the distance between neighboring sites.

We then expand the action for small $\vec{L}$, take the continuum limit and integrate out the average magnetization fluctuations of $\vec{L}$. An action that depends solely on the coarse-grained $SO(3)$-field $\hat{R}(x,\tau)$ emerges,
\begin{eqnarray}
	S &=& -\frac{\rho_s}{2} \int d^2x \int_0^\beta d\tau \left(
	\frac{1}{c^2} \Tr \left[ (\hat{R}^{-1} \partial_\tau \hat{R})^2 \right] 
	\right. \nonumber \\ && \left.
		+ \Tr \left[ \hat{P} (\hat{R}^{-1} \vec{\nabla} \hat{R})^2 \right]
	\right).
	\label{SO3action}
\end{eqnarray}
where $\hat{P}_{\alpha \beta} = \frac{3}{2} \sum_{i=A,B,C} n_{i \alpha} n_{i \beta}$ is the projection operator onto the plane of the magnetic order of the reference state. This projection causes the spin waves to have different velocities depending on whether their polarization is in the plane of the order or out-of-plane.

The nearest-neighbor triangular Heisenberg model gives rise to the $SO(3)$ non-linear sigma model of Eqn. (\ref{SO3action}). In general, based on symmetry argments,\cite{Sachdev2011qpt..book} an $SO(3)$ model can also contain terms of the form $\Tr \left[ \hat{P} (\hat{R}^{-1} \partial_\tau \hat{R})^2 \right]$ and $\Tr \left[ (\hat{R}^{-1} \vec{\nabla} \hat{R})^2 \right]$. Let us now discuss these terms in full generality.

\subsection{Charts on $SO(3)$}

It seems that the model Eqn. (\ref{SO3action}) depends on the reference state, which should not be the case for a symmetric theory. However, by choosing the appropriate chart on $SO(3)$ we can reveal the correct anisotropy structure.

The group of rotations in three dimensions has three degrees of freedom, which can be parametrized in various ways. Starting with the standard basis for the corresponding Lie algebra $so(3)$ of infinitesimal rotations,
\begin{equation}
	T^1 = {\small \begin{pmatrix} 
		0 & 0 & 0 \\
		0 & 0 & -1 \\ 
		0 & 1 & 0 \end{pmatrix}}, \,
	T^2 = {\small \begin{pmatrix} 
		0 & 0 & 1 \\
		0 & 0 & 0 \\ 
		-1 & 0 & 0 \end{pmatrix}}, \,
	T^3 = {\small \begin{pmatrix} 
		0 & -1 & 0 \\
		1 & 0 & 0 \\ 
		0 & 0 & 0 \end{pmatrix}},
\end{equation}
we can parametrize elements of $SO(3)$ using the exponential map
\begin{equation}
	\hat{R} = \hat{R} (\vec{\omega}) = e^{\vec{\omega} \cdot \vec{T}}
\end{equation}
where $|\vec{\omega}| \leq \pi$. The derivatives $\hat{R}^{-1} \partial_\mu \hat{R}$ in the model Eqn. (\ref{SO3action}) can be expanded for small rotations $|\vec{\omega}| \ll 1$,
\begin{eqnarray}
	\Tr (\hat{R}^{-1} \partial_\mu \hat{R})^2
		&=& -2 \frac{4 \sin^2 \frac{|\vec{\omega}|}{2}}{|\vec{\omega}|^2} (\partial_\mu \vec{\omega})^2
		\nonumber \\ && 
			- \frac{2}{|\vec{\omega}|^2}\left(|\vec{\omega}|^2-4\sin^2 \frac{|\vec{\omega}|}{2} \right) (\partial_\mu |\vec{\omega}|)^2 \nonumber \\
		& \approx &
		-2 (\partial_\mu \vec{\omega})^2 + \ldots
\end{eqnarray}
where in the last line we have expanded for small $\omega$ up to fourth order. Notice that small $|\vec{\omega}| \ll 1$ corresponds to small rotations. Similarly, the anisotropic part can be expanded as $\Tr \hat{P} (R^{-1} \partial_\mu R)^2 \approx - (\partial_\mu \omega_x)^2- (\partial_\mu \omega_y)^2- 2(\partial_\mu \omega_z)^2$.

A more insightful parametrization is through the rotation quaternions: four real numbers $a_i$, $i=0, \ldots 3$, with the constraint $a_i^2 = 1$. A rotation matrix as a function of quaternions is
\begin{equation}
	R_{ij} = \delta_{ij} \left(a_0^2 - \sum_{k=1}^3 a_k a_k \right)+ 2 a_i a_j + 2 \sum_{k=1}^3 \epsilon_{ijk} a_0 a_k
\end{equation}
and the integration measure is the same as for the 3-sphere,
\begin{equation}
	\int_{SO(3)} dR = \frac{1}{2\pi^2} \int_{|\vec{a}|=1} d^4 \vec{a}.
\end{equation}
The isotropic term in the nonlinear sigma model now becomes
\begin{equation}
	\Tr (\hat{R}^{-1} \partial_\mu \hat{R})^2
	= - 8 (\partial_\mu \vec{a})^2.
\end{equation}
The anisotropic term becomes quite elegant in this notation. The anisotropy direction should, of course, depend on the average magnetization, but the model action has to be symmetric. This can be done by introducing a smooth vector field on the $3$-sphere, $\vec{v}(\vec{a}) = \gamma^1 \vec{a}$, that defines the local anisotropy direction. With the specific choice of vector field direction,
\begin{equation}
	\gamma^1 = {\small \begin{pmatrix} 
		0 & 0 & 0 & 1 \\
		0 & 0 & 1 & 0 \\ 
		0 & -1 & 0 & 0 \\
		-1 & 0 & 0 & 0\end{pmatrix}}
\end{equation}
we can use this \emph{anisotropy vector field} to write
\begin{equation}
	\Tr (1-P) (R^{-1} \partial_\mu R)^2 
	= 4 (\vec{a} \gamma^1 \partial_\mu \vec{a})^2. 
\end{equation}
Therefore, the most general $SO(3)$ nonlinear sigma model action is
\begin{widetext}
\begin{equation}
	S = \frac{1}{2} \int d^2x \int_0^\beta d\tau \, \left(
		\chi_\parallel (\partial_\tau \vec{a})^2 
		+ \rho_{\parallel} (\vec{\nabla} \vec{a})^2
			+ ( \chi_\perp -  \chi_\parallel ) (\vec{a} \gamma^1 \partial_\tau \vec{a})^2
			+ (\rho_\perp - \rho_\parallel) (\vec{a} \gamma^1 \vec{\nabla} \vec{a})^2
	\right)
	\label{FullSO3action}
\end{equation}
\end{widetext}
where $\chi$ is the spin susceptibility and $\rho$ is the spin stiffness. Renormalization group studies\cite{Azaria92,Chubukov1994PhRvL,Chubukov1994NuPhB} argue that any anisotropy (that is $\rho_\perp \neq \rho_\parallel$ or $\chi_\perp \neq \chi_\parallel$) is dangerously irrelevant close to the quantum phase transition to the disordered phase. Furthermore, a Holstein-Primakoff (large $S$) expansion has been performed around the coplanar ordered state,\cite{Chubukov:5ZNJ-9uZ, Chubukov:liyMxAMw,Lechimant95} finding anisotropy in both the susceptibility as well as in the spin stiffness. 

Notice that the quaternion chart is equivalent to the chart of two complex numbers $z_i$ with $|z_i|^2 = 1$ to represent rotations.\cite{Chubukov1994PhRvL,Sachdev2011qpt..book}

\section{Entanglement spectrum}
\label{SecEntg}
Having introduced coplanar antiferromagnetism and the most general action that describes it, we proceed by analytically computing the entanglement spectrum of the ground state of Eqn. (\ref{FullSO3action}). We thereby follow the approach of Ref. [\onlinecite{Metlitski:akSbyvLD}], which uses the nonlinear sigma model to derive the entanglement spectrum of a collinear magnet.

The entanglement entropy has been computed numerically by Kolley et al. for the triangular lattice\cite{Kolley2013arXiv1307.6592K} as well as for the Kagome lattice\cite{Kolley2014arXiv1410.7911K}. We find that our analytical results are consistent with their numerical results.

\subsection{Ground state wavefunction}

First we need to find the ground state wavefunction of the model Eqn. (\ref{FullSO3action}). Therefore we split the $\vec{k}=0$ component from the spin wave fluctuations,
\begin{equation}
	\vec{a}(\vec{x},\tau) = 
		\vec{n}_0(\tau) \sqrt{ 1 - \pi_i (\vec{x},\tau)^2/\rho_i}
			+ \vec{e}_i \pi_i(\vec{x},\tau) / \sqrt{\rho_i}
\end{equation}
where $\vec{e}_i$ with $i=1, 2, 3$ are unit vectors orthogonal to $\vec{n_0} (\tau)$. The anisotropy is parametrized by $\rho_i = \rho_\parallel$ for $i=1,2$ and $\rho_\perp$ for $i=3$. Notice that the 3-direction is defined by the anisotropy vector field $\gamma^1 \vec{n}_0$. In order to keep the same number of degrees of freedom, the $\pi_i$ fields cannot have zero momentum, which means $\int d^2x \pi_i(\vec{x},\tau) = 0$. The anisotropic term expanded for small $\pi_i$ yields an extra term for the 3rd spin wave field,
\begin{equation}
	(\vec{a} \gamma^1 \partial_\mu \vec{a})^2 = (\partial_\mu \pi_3(\vec{x},\tau))^2 + \ldots
\end{equation}
and so the action becomes, up to quadratic order,
\begin{eqnarray}
	S &= &\frac{V \chi_\parallel}{2 } \int_0^\beta d\tau (\partial_\tau \vec{n}_0)^2
	\nonumber \\ && 
		+ \frac{1}{2} \int d^2x \int_0^\beta d\tau \left( \frac{1}{c_i^2} (\partial_\tau \pi_i)^2 + (\vec{\nabla} \pi_i)^2 \right)
\end{eqnarray}
where $c_i^2 = \rho_i / \chi_i$ represents the spin wave velocity squared. 

We will now transform back to the Hamiltonian formalism. The Hamiltonian associated with the zero-momentum part $\vec{n}_0(\tau)$ is that of a free particle on a $3$-sphere,
\begin{equation}
	H_t = \frac{L_{\alpha \beta}^2}{2\chi_\parallel V}.
	\label{N0Component}
\end{equation}
The ground state has zero angular momentum, which is a singlet, and has a wavefunction that is constant on the 3-sphere.

The Hamiltonian for the spin-wave part, that is $\vec{k} \neq 0$, is
\begin{equation}
	H_{sw} = \sum_{\vec{k}\neq0, i} \epsilon^i_{\vec{k}} b^\dagger_{\vec{k} i} b_{\vec{k} i}
\end{equation}
where the dispersion $\epsilon^i_{\vec{k}} = c_\parallel |\vec{k}|$ for the inplane modes $i=1,2$, and $\epsilon^3_{\vec{k}} = c_\perp |\vec{k}|$ for the out-of-plane mode. Bear in mind that $c_i = \sqrt{ \rho_i / \chi_i}$. The static propagator that should be reproduced by the ground state wavefunction is
\begin{equation}
	\langle \pi_i(\vec{x}) \pi_j (\vec{x}) \rangle
		= D_{ij} (\vec{x},\vec{y}) 
		= \delta_{ij} \frac{1}{V} \sum_{\vec{k} \neq 0} 
		\frac{1}{2 \epsilon^i_{\vec{k}}}
		e^{i \vec{k}\cdot(\vec{x}-\vec{y})}
\end{equation}
and hence, with the inverse of $D$ defined as
\begin{equation}
	\int d^2z Q_{ij} (\vec{x}, \vec{z}) D_{jk} (\vec{z}, \vec{y})
		= \left( \delta^2 (\vec{x}-\vec{y}) - \frac{1}{V} \right) \delta_{ik}
\end{equation}
we find the ground state wavefunction of the spin-wave part is
\begin{equation}
	\psi[\pi_i] \propto
		\exp \left( -\frac{1}{4} \int d^2x d^2y \pi_i (\vec{x}) Q_{ij} (\vec{x}, \vec{y}) \pi_j (\vec{y}) \right)
\end{equation}
The contribution from spin-waves is anisotropic, depending on the direction of the magnetic order, characterised by $\vec{n}_0 \in SO(3)$. One must thus bear in mind that the degrees of freedom $\pi_i(\vec{x})$ are associated with a specific choice of the ordering direction. In particular, if we choose the global rotation $\vec{n}_0 = (1,0,0,0)$, the corresponding $SO(3)$ element is the identity, and hence the order the same as in the reference state: in the spin $xy$-plane. In this case, we can identify the 3rd spin wave field $\pi^3 (\vec{x})$ with the actual $z$-component of the spin. 

For further convenience, we introduce the inverse correlator $Q(\vec{x},\vec{y})$ without dependence on the spin-wave velocity,
\begin{equation}
	Q(\vec{x},\vec{y}) = \sum_{\vec{k} \neq 0 } 2 |\vec{k}| e^{i \vec{k} \cdot (\vec{x} - \vec{y})}.
	\label{Qdef}
\end{equation}
and the spin wave anisotropy matrix
\begin{equation}
	\hat{c} = {\small \begin{pmatrix}
		c_\parallel & 0 & 0 \\
		0 & c_\parallel & 0 \\
		0 & 0 & c_\perp
		\end{pmatrix}}
		= c_\parallel \hat{1} + (c_\perp - c_\parallel) \hat{P}^3.
	\label{cdef}
\end{equation}
where $\hat{P}^3$ projects onto the third axis. With these definitions, the ground state wavefunction for the $SO(3)$ field is 
\begin{equation}
	\psi[\vec{n}_0, \pi_i] \propto
		\exp \left( -\frac{1}{4} \int d^2x d^2y \pi_i (\vec{x}) c_{ij} Q(\vec{x}, \vec{y}) \pi_j (\vec{y}) \right).
	\label{GSWF}
\end{equation}

\subsection{Reduced Density Matrix}

Our next step is to compute the reduced density matrix $\rho_A$. The density matrix of the total system is defined as $\rho(\vec{a},\vec{a}') = \psi[\vec{a}] \psi^*[\vec{a}']$. In order to obtain the reduced density matrix, the total system is split into region $A$ and region $B$, with volume $V_A$ and $V_B$, respectively. We will give the degrees of freedom on each region a corresponding subscript, so $\vec{a}(\vec{x}) = \vec{a}_A(\vec{x}) + \vec{a}_B(\vec{x}).$ In order to get the reduced density matrix on region $A$, that is $\rho_A = \Tr_B \rho$, we must identify the two fields on region $B$, that is $\vec{a}_B(\vec{x}) = \vec{a}_B'(\vec{x})$. Because the ground state wavefunction is Gaussian and thus deviations from the average magnetic order are exponentially suppressed, we can expand $\vec{a}$ and $\vec{a}'$ both around a common choice of directions, and subsequently integrating out $\pi_B$. Later, we need to restore the broken symmetry of choosing a direction around which to expand.

Let us expand the fields around the north pole of the $3$-sphere,
\begin{eqnarray}
	a^0(\vec{x}) &= &
		\sqrt{ 1 - \pi_i (\vec{x})^2/\rho_i} 
		\label{NP1} \\
	a^i (\vec{x}) & = &
		\pi_i(\vec{x}) / \sqrt{\rho_i}.
		\label{NP2}
\end{eqnarray}
The ground state wavefunction is still Eqn. (\ref{GSWF}), but now the fields $\pi_i(\vec{x})$ are unconstrained and can have a zero momentum component. In fact, the magnetic order of a field configuration $\pi^i(\vec{x})$ is given by the $\vec{k}=0$ component,
\begin{equation}
	\vec{n}_0 = \vec{e}_0 + \frac{1}{V} \int d^2 x \pi^i(\vec{x}) \vec{e}_i / \sqrt{\rho_i}
\end{equation}
up to first order in $\pi^i(\vec{x})$, where $\vec{e}_0 = (1,0,0,0)$, and $\vec{e}_i \cdot \vec{e}_j = \delta_{ij}$ form a four-dimensional orthogonal basis. Notice that since $\int d^2x Q(x,y) = 0$, the inclusion of the zeroeth momentum component does not change the ground state wavefunction.

It does allow us, however, to relate $\pi^i_B (\vec{x}) = {\pi^i_B}' (\vec{x})$. We introduce the short-hand notation
\begin{equation}
	\pi_A \hat{Q}_{AB} \pi_B
		= \int_A d^2x \int_B d^2y \, \pi_A^i Q_{ij}(\vec{x}, \vec{y}) \pi_B^j,
\end{equation}
where $\hat{Q}_{BB}$ is the matrix operator where the first and second coordinate are both in region $B$, and so forth for similar expressions. The reduced density matrix now becomes
\begin{widetext}
\begin{eqnarray}
	\rho_{A} & \propto &
		\int \mathcal{D}\pi_B \, \exp \left[
		-\frac{1}{4} \left(
		\pi_A \hat{Q}_{AA} \pi_A + \pi_A' \hat{Q}_{AA} \pi_A' 
		+ (\pi_A + \pi_A') \hat{Q}_{AB} \pi_B + \pi_B \hat{Q}_{BA} (\pi_A+\pi_A')
		+ 2 \pi_B \hat{Q}_{BB} \pi_B
		\right) \right] \nonumber \\
	&\propto& \exp \left[
		-\frac{1}{8} \left(
		2 \pi_A \hat{Q}_{AA} \pi_A + 2 \pi_A' \hat{Q}_{AA} \pi_A' 
		- (\pi_A + \pi_A') \hat{Q}_{AB} \hat{Q}_{BB}^{-1} \hat{Q}_{BA} (\pi_A+\pi_A')
		\right) \right].
\end{eqnarray}
where $\int_B d^2z \hat{Q}_{BB}^{-1}(\vec{x},\vec{z}) \hat{Q}_{BB} (\vec{z},\vec{y}) = \delta^2(\vec{x}-\vec{y}) \hat{1}$. Notice that the anisotropy matrix-structure of $\hat{Q}_{AA}$ and $\hat{Q}_{AB} \hat{Q}_{BB}^{-1} \hat{Q}_{BA}$ are the same, namely that of the spin-wave matrix anisotropy matrix Eqn. (\ref{cdef}). Using the scalar definition of $Q(\vec{x},\vec{y})$ of Eqn. (\ref{Qdef}), it follows that
\begin{equation}
	\rho_A[\pi_A, \pi_A'] \propto
	\exp \left[
		-\frac{1}{8} \left(
		2 \pi_A \hat{c} Q_{AA} \pi_A + 2 \pi_A' \hat{c} Q_{AA} \pi_A' 
		- (\pi_A + \pi_A') \hat{c}Q_{AB} Q_{BB}^{-1} Q_{BA} (\pi_A+\pi_A')
		\right) \right]
\end{equation}
 We can write this result in terms of the sum and difference of $\pi_A$ and $\pi_A'$,
\begin{equation}
	\rho_A[\pi_A, \pi_A'] \propto
	\exp \left[
		-\frac{1}{8} (\pi_A - \pi_A') \hat{c} Q_{AA} (\pi_A-\pi_A')
		-\frac{1}{8}
		(\pi_A + \pi_A') \hat{c}\left( Q_{AA} - Q_{AB} Q_{BB}^{-1} Q_{BA} \right)(\pi_A+\pi_A')
	\right].
	\label{SpinWaveRhoA}
\end{equation}
\end{widetext}
The anisotropy in this result is relative to the expansion point we chose in Eqns. (\ref{NP1})-(\ref{NP2}). In order to restore the symmetry we split off the $\vec{k}=0$ component of the $\pi_A$-fields from the $\vec{k}\neq0$ components. That is achieved by a Fourier transform the fields and $Q$-functions. The difference term in the reduced density matrix reads
\begin{equation}
	\exp \left[ - \frac{1}{8} V_A^2 \sum_{\vec{k}} (\pi_A - \pi_A')^i (\vec{k}) c_{ij}
		Q(\vec{k}) (\pi_A - \pi_A')^j (-\vec{k}) \right].
\end{equation}
Recall that in order to derive this expression, we assumed that $\pi_A$ and $\pi_A'$ are close to each other, so that the local anisotropy vector field can be approximated by $\frac{1}{2} \gamma^1 (\vec{n} + \vec{n}')$. The anisotropic part of the wavefunction, which defines the orthogonal direction, can thus be written as
\begin{equation}
	\left( \frac{1}{2} (\vec{n} + \vec{n}')\gamma^1   (\vec{n} - \vec{n}') \right)^2
	= (\vec{n} \gamma^1 \vec{n}')^2
\end{equation}
Notice also that the $\vec{k}=0$ component of $\left( Q_{AA} - Q_{AB} Q_{BB}^{-1} Q_{BA} \right)$ is zero\cite{Metlitski:akSbyvLD} so that doesn't contribute to the $\vec{k}=0$ part of the reduced density matrix, which becomes
\begin{eqnarray}
	\rho_A [ \vec{n}_{A,0}, \vec{n}_{A,0}']&
	\nonumber \\ 
	\propto & 
		\exp \left[ -\frac{I}{2} \rho_\parallel c_\parallel (\vec{n}_{A,0}-\vec{n}'_{A,0})^2 \right]
			\label{ReducedN0} \\
	& \times	\exp \left[ - \frac{I}{2} (\rho_\perp c_\perp - \rho_\parallel c_\parallel)\left( \vec{n}_{A,0} \gamma^1 \vec{n}_{A,0}' \right)^2 
	\right] \nonumber
\end{eqnarray}
where the effective moment of inertia is given by
\begin{equation}
	I = \frac{1}{4} \int_A d^2x d^2y Q(\vec{x}, \vec{y}).
	\label{MomentofIntertia}
\end{equation}

\subsection{The Tower of States}
We have come to the centerpiece of this paper: the entanglement spectrum of the $\vec{k}=0$ component of the reduced density matrix, Eqn. (\ref{ReducedN0}), displays the tower of states.

To prove that, we need to find the \emph{entanglement Hamiltonian} $H^E_{tos}$ such that
\begin{equation}
	\rho_A [ \vec{n}_{A,0}, \vec{n}_{A,0}'] = \langle \vec{n}_{A,0} | e^{-H^E_{tos}} | \vec{n}_{A,0}'\rangle.
\end{equation} 
The density matrix Eqn. (\ref{ReducedN0}) describes the motion of a particle on a $3$-sphere. The dynamics of a particle constraint to move on a circle should be expressed in terms of the angular momentum,\cite{Sachdev2011qpt..book}
\begin{equation}
	L_{\alpha \beta} = n_\alpha p_\beta - n_\beta p_\alpha 
\end{equation}
with the usual commutation relation $[n_\alpha, p_\beta]=i\delta_{\alpha\beta}$. On the $3$-sphere, the total angular momentum squared $\sum_{\alpha < \beta} L^2_{\alpha \beta}$ is equal to the Laplacian $\vec{p}^2$ so that for large $I$ we have
\begin{eqnarray}
	\langle \vec{n} | e^{-\frac{L^2}{2I}} | \vec{n'} \rangle
	& \approx &
		\langle \vec{n} | e^{-\frac{p^2}{2I}} | \vec{n'} \rangle 
		\nonumber \\
	&=& \int \frac{d^3p}{(2\pi)^3} e^{i \vec{p} \cdot (\vec{n} - \vec{n}') - \frac{p^2}{2I}} \nonumber \\
	& = & \left(\frac{I}{2\pi}\right)^{3/2} \exp \left\{ - \frac{I}{2} (\vec{n} - \vec{n}')^2 \right\}.
	\label{LaplaceOnASphere}
\end{eqnarray}
Using the anisotropy vector field $\gamma^1$, it follows that
\begin{equation}
	(\vec{n} \gamma^1 \vec{p})^2 
	= \left( \frac{1}{2} \sum_{\alpha \beta} L_{\alpha \beta} \gamma_{\alpha \beta}^1 \right)^2
		= \frac{1}{4} (\Tr \; L \gamma^1)^2.
\end{equation}
and similar to Eqn. (\ref{LaplaceOnASphere}) we can find the matrix elements of the ansatz
\begin{equation}
	H^E = \frac{1}{2} a \hat{p}^2 + \frac{1}{2} b (\hat{n} \gamma^1 \hat{p})^2
\end{equation}
so that, using a saddle-point integration and the fact that $(\gamma^1)^2=-1$,
\begin{equation}
	\langle y | e^{-H^E} | x \rangle
		\sim \exp \left[ - \frac{1}{2a} (\vec{x} - \vec{y})^2
		- \frac{1}{2} \frac{b/a}{a+b} (\vec{y} \gamma^1 \vec{x})^2 \right].
\end{equation}
Equating this result with the reduced density matrix, Eqn. (\ref{ReducedN0}) gives the entanglement Hamiltonian
\begin{equation}
	H^E_{tos} = \frac{1}{2I \rho_\parallel c_\parallel} \sum_{\alpha > \beta} L_{\alpha \beta}^2
		+ \frac{(\rho_\perp c_\perp - \rho_\parallel c_\parallel)}{8 I \rho_\parallel c_\parallel (2 \rho_\parallel c_\parallel - \rho_\perp c_\perp)} (\Tr \, L \gamma^1 )^2.
	\label{EntanglementH}
\end{equation}
Finally, we are in a position to find the low-lying states of the entanglement spectrum. For that we need to find the representation of the angular momentum algebra of the particle on the $3$-sphere. This algebra is isomorphic to $O(3) \times O(3)$, which can be made explicit by defining\cite{VanIsacker:1991if,Wybourne:1974uv}
\begin{eqnarray}
	J_i & = & \frac{1}{2} \left( \frac{1}{2} \epsilon_{ijk} L_{jk} + L_{4i} \right) \\
	N_i & = & \frac{1}{2} \left( \frac{1}{2} \epsilon_{ijk} L_{jk} - L_{4i} \right)
\end{eqnarray}
such that their commutation relations are
\begin{eqnarray}
	[ J_i , J_j] & = & i \epsilon_{ijk} J_k \\
	\left[ N_i , N_j \right] & = & i \epsilon_{ijk} N_k \\
	\left[ J_i , N_j \right] & = & 0.
\end{eqnarray}
Note, however, that $\vec{J}^2 = \vec{N}^2 = \frac{1}{4} L_{\alpha \beta}^2$, so that the representations of $O(4)$ can be labeled by a total angular momentum $j$. States in this representation are labelled by
\begin{equation}
	| \{ j \} m_1 m_2 \rangle
\end{equation}
where $m_i = -j , -j+1, \ldots j$ are the quantum numbers for each $O(3)$ subalgebra. For each value of the total angular momentum there are $(2j+1)^2$ states. Note also that in this representation the anisotropy term becomes relatively simple,
\begin{equation}
	\frac{1}{4} (\Tr L \gamma^1)^2 = (N_1)^2.
\end{equation}
The eigenvalues of the entanglement Hamiltonian Eqn. (\ref{EntanglementH}) are thus
\begin{eqnarray}
	E_{j,m_1,m_2} &=& \frac{2}{I \rho_\parallel c_\parallel} \left( j (j+1) \right) \nonumber \\
	&&
		+ \frac{(\rho_\perp c_\perp - \rho_\parallel c_\parallel)}{2 I \rho_\parallel c_\parallel (2 \rho_\parallel c_\parallel - \rho_\perp c_\perp)} (m_2)^2
	\label{EntanglementSpectrum}
\end{eqnarray}
This corresponds to the degeneracies observed in Ref. \cite{Kolley2013arXiv1307.6592K}. For each value of total spin $j$, there are $(2j+1)^2$ states in the entanglement spectrum, split into $2j+1$ states with different value of $m_2$. The structure for each value of $j$, which is shown in Fig. 2 and 3 of Ref. \cite{Kolley2013arXiv1307.6592K}. The fact that for integer $j$ the highest energy state is single degenerate, suggests that this highest energy state has $m_2=0$. Consequently, the prefactor in front of the $(N_1)^2$ term is negative for the model they studied.

We can compare our results with the square lattice antiferromagnet, where the order parameter is an element of the $2$-sphere symmetry. In that case the entanglement spectrum has a $(2S+1)$ degeneracy\cite{Kallin2011PhRvB..84p5134K}. Only in the case of the triangular and Kagom\'{e} antiferromagnets, where there is an $SO(3)$ \emph{coplanar} instead of a collinear $S_2$ order parameter, we find a $(2S+1)^2$ degeneracy.

The attentive reader might ask: what happened to the $\vec{k}\neq0$ components of the reduced density matrix? Here we can be brief, as this part has the shape of the density matrix of an harmonic oscillator,
\begin{widetext}
\begin{equation}
	\rho_A [\pi,\pi'] \propto
		\exp \left[ - \frac{1}{4} (\pi-\pi') \hat{M}_1 (\pi-\pi')
		- \frac{1}{4} (\pi+\pi') \hat{M}_2 (\pi + \pi') \right].
\end{equation}
\end{widetext}
We can thus directly apply the results from Ref. [\onlinecite{Metlitski:akSbyvLD}] to find that the entanglement Hamiltonian for the spin wave spectrum is just that of free Goldstone bosons, the difference only being in the anisotropy of the spin wave velocities.

The full entanglement Hamiltonian is thus given by the tower of states contribution of Eqn. (\ref{EntanglementH}) and the spin-wave part,
\begin{equation}
	H^E_{tos} = \frac{L_{\alpha \beta}^2}{2I_1}
		+ \frac{(\Tr \, L \gamma^1 )^2}{2 I_2} 
	\label{EntanglementHFull}
	+ \sum_{\epsilon,\alpha} \epsilon_{\alpha} a^\dagger_{\epsilon,\alpha} a_{\epsilon,\alpha},
\end{equation}
where we have, for compact notation, absorbed the $\rho$ and $c$-dependence into the parameters $I_1$ and $I_2$, and $\alpha$ runs over the three different spin waves with energies $\epsilon_\alpha$.

How do the parameters $I_1, I_2$ and $\epsilon_\alpha$ depend on the size of the subregion? Here we follow the results from Ref. [\onlinecite{Metlitski:akSbyvLD}] again. Consider a two-dimensional torus divided into two cylinders $A$ and $B$ by making cuts at $x=0$ and $x=\ell$. On such a symmetry, the lowest energy eigenstate for the spin-wave spectrum scales at $\epsilon \sim \left( \log \ell/a \right)^{-1}$, where $a$ is the UV cut-off. The moment of inertia parameters Eqn. (\ref{MomentofIntertia}), on the other hand, scale as 
\begin{equation}
	I_i \sim \ell \log (\ell/a)
\end{equation}
which makes the gap to the first excited tower state an order of magnitude $\ell$ smaller. Therefore, the tower of states entanglement spectrum is separated from the spin wave entanglement spectrum.

In conclusion, the low energy entanglement spectrum has the same structure as the tower of states. Therefore, the degeneracies of the entanglement spectrum allow for a determination of the symmetry of the relevant order parameter, similar to the idea proposed that the entanglement spectrum allows for a determination of topological order.\cite{Li:2008cg}

\subsection{Entanglement entropies}

Finally, let us mention the consequences of the $SO(3)$ order parameter to the entanglement entropy. The Von Neumann entanglement entropy is defined as
\begin{equation}
	S_{E} = -\Tr \left( \rho_A \log \rho_A \right)
\end{equation}
or can be equivalently found as the limit of Renyi entropies
\begin{equation}
	S_{n} = -\frac{1}{n-1} \Tr \left( \rho_A^n \right)
\end{equation}
such that $\lim_{n\rightarrow1} S_n = S_{E}$. 

It turns out to be quite undemanding to compute the Renyi entropies. Let us therefore focus first on the tower of states part. Starting from the reduced density matrix expressed in terms of the $\pi$-vectors, we can get higher powers of the reduced density matrix $\rho_A^n[\pi,\pi'']$ by Gaussian integration. Recall that the relevant part of the reduced density matrix is
\begin{equation}
	\rho_A[\pi_0,\pi_0']  \sim 
	\exp \left[ - \frac{1}{2} I (\pi_0 - \pi_0')^i c_{ij} (\pi_0 - \pi_0')^j \right].
\end{equation}
where $I \sim \ell \log \ell/a $. The reduced density matrix squared is
\begin{widetext}
\begin{eqnarray}
	\rho_A^2[\pi_0,\pi_0''] & \sim &
	\int \frac{d^3 \pi_0'}{\sqrt{\det \hat{\rho}}}
	\exp \left[ - \frac{1}{2} I (\pi_0 - \pi_0')^i c_{ij} (\pi_0 - \pi_0')^j
		- \frac{1}{2} I (\pi_0' - \pi_0'')^i c_{ij} (\pi_0' - \pi_0'')^j \right] \\
	&=&
	\int \frac{d^3 \delta n}{\sqrt{\det \hat{\rho}}}
	\exp \left[ - I \delta n^i c_{ij} \delta n^j
		+ I \delta n^i c_{ij} (\pi_0 - \pi_0'')^j 
		- \frac{1}{2} I (\pi_0 - \pi_0'')^i c_{ij} (\pi_0 - \pi_0'')^j\right] \\
	&=&
		\left( \frac{2\pi}{2I} \right)^{3/2} (\det \hat{c} \hat{\rho})^{-1/2} 
		\exp \left[ - \frac{1}{2} I (\pi_0 - \pi_0'')^i c_{ij} (\pi_0 - \pi_0'')^j \right].
\end{eqnarray}
Notice that we included the spin stiffness as part of the measure on the integration over $\pi$. In the second line, we defined $\delta n = \pi_0 - \pi_0'$. This integration technique can be continued to find the $n$-th power,
\begin{equation}
	\rho_A^n[\pi_0,\pi_0'']  \sim 
		n^{-3/2}
		\left( \frac{2\pi}{I} \right)^{3(n-1)/2} (\det \hat{c} \hat{\rho})^{-(n-1)/2} 
		\exp \left[ - \frac{I}{2n} (\pi_0 - \pi_0'')^i c_{ij} (\pi_0 - \pi_0'')^j \right].		
\end{equation}
\end{widetext}
The Renyi entropy associated with this result is
\begin{equation}
	S_n
		= \frac{3}{2} \log \left( \frac{I}{2\pi} \right) + \frac{3}{2} \frac{\log n}{n-1} + \frac{1}{2} \log \det \hat{c} \hat{\rho}.
\end{equation}
The entanglement entropy of the tower of states becomes
\begin{eqnarray}
	S_E &=& \frac{3}{2} \log \left( \frac{I}{2\pi} \right) + \frac{3}{2} + \frac{1}{2} \log \det \hat{c} \hat{\rho} \\
	 &\sim &\frac{3}{2} \log \ell + \mathrm{const.}
\end{eqnarray}
The tower of states thus introduces a logarithmic term in the entanglement entropy. The large-size of this term is, to leading order in $\ell$, independent of the anisotropy introduced by the coplanar order. In fact, it confirms the picture that the logarithmic correction in symmetry broken systems scales as $\frac{N-1}{2} \log \ell$ where $N-1$ is the number of Goldstone modes.

The spin wave part will contribute to the entanglement entropy in a similar fashion to the collinear results, which yields an area law $S_E \sim \ell$ with a nonuniversal prefactor.

\section{Conclusion}

Using an $SO(3)$ nonlinear sigma model, we have shown that the ground state entanglement spectrum of a coplanar antiferromagnet displays the 'tower of states' structure. States in this tower can be labelled by total spin $j$ and have $(2j+1)^2$ degeneracy which is lifted by the spin-wave anisotropy. Consistent with earlier results for the collinear antiferromagnet, the entanglement entropy counts the number of Goldstone modes of the broken symmetry state.

These results point towards interesting universal behavior. Will a system that undergoes spontaneous symmetry breaking always hide this in its finite size ground state entanglement? Clearly, ferromagnetic ground states do not exhibit this behavior as they are known to be exact product states. For the antiferromagnets studied here and in Ref. [\onlinecite{Metlitski:akSbyvLD}], we already knew that there was a tower of states associated with the symmetry breaking. If one would be able to prove in general, that systems that break a symmetry in the thermodynamic limit\footnote{With the exception of ferromagnet-like systems where the order parameter commutes with the Hamiltonian.} should have a tower of states in their entanglement spectrum and a logarithmic correction to the entanglement entropy, then such a result would be invaluable to search and identify yet unknown broken symmetry phases of any Hamiltonian.

\acknowledgments
The author thanks Tarun Grover, Max Metlitski and Aron Beekman for discussions, and Dominique Mouhanna and Philippe Lecheminant for pointing out relevant references. L.R. was supported by the Dutch Science Foundation (NWO) through a Rubicon grant.


\end{document}